\documentclass[prl,twocolumn,aps,a4paper,superscriptaddress,showpacs]{revtex4}
\usepackage{amsmath,latexsym,amsfonts}
\usepackage[dvips]{graphicx}
\begin{document}
\title{Non Abelian gauge symmetries induced by the unobservability of extra-dimensions in a Kaluza-Klein approach}

\author{Francesco Cianfrani}
\email{francesco.cianfrani@icra.it}
\author{Giovanni Montani}
\email{montani@icra.it} 
\affiliation{Dipartimento di Fisica Universit\`a di Roma ``La Sapienza''}
\affiliation{ICRA---International Center for Relativistic Astrophysics  
c/o Dipartimento di Fisica (G9) Universit\`a di Roma ``La Sapienza'',
Piazza A.Moro 5 00185 Rome, Italy}

\today

\begin{abstract}
In this work we deal with the extension of the Kaluza-Klein approach to a non-Abelian gauge theory; we show how we need to consider the link between the n-dimensional model and a four-dimensional observer physics, in order to reproduce fields equations and gauge transformations in the four-dimensional picture. More precisely, in fields equations any dependence on extra-coordinates is canceled out by an integration, as consequence of the unobservability of extra-dimensions. Thus, by virtue of this extra-dimensions unobservability, we are able to recast the multidimensional Einstein equations into the four-dimensional Einstein-Yang-Mills ones, as well as all the right gauge transformations of  fields are induced. The same analysis is performed for the Dirac equation describing the dynamics of the matter fields and, again, the gauge coupling with Yang-Mills fields are inferred from the multidimensional free fields theory, together with the proper spinors transformations.
\end{abstract}



\maketitle

\section{Introduction}
In modern physics we have to face with a strong difference between the formalisms that describe interactions; in fact while we interpret the gravitational field as a geometric propriety of space-time, other interactions rely on the notion of gauge theories, i.e. models based on the invariance of the theory under a local group of transformations involving fields. One of the most relevant issue of the gauge approach is the one in which gauge bosons naturally arise to preserve the invariance.\\
The first attempt to get a geometrical unified picture of general relativity with a $U(1)$ Abelian gauge theory (the electrodynamics) has to be remanded to Kaluza \cite{1} and this approach relays on the introduction of an extra-dimension; in a modern point of view, in his work gauge transformations are reproduced by translations on the extra-coordinate and, from the 5-dimensional curvature splitting, Einstein-Hilbert-Maxwell action outcomes. The role of the extra-dimensional space and its unobservability, as related to the compactification and to the quantum uncertainty, was clarified by Klein \cite{2}, \cite{3}, who also gave an explanation for the \emph{cylindricity condition}, i.e. the fundamental hypothesis addressed by the Kaluza theory, implying that physical quantities do not have a dependence on the extra-coordinate.\\      
The Kaluza-Klein approach can be extended to more complex group \cite{Ke68}, \cite{MKKT}(for a review see \cite{OW97}, for the problems related to the application to physical interactions see \cite{W81}, \cite{W83}) just considering a more complex extra-dimensional homogeneous \cite{Lan} space with Killing vectors algebra reproducing the Lie one of the gauge group we wish to geometrize; however this extension is not straightforward, because we do not reproduce the correct transformation law for non-Abelian gauge bosons and we do not have an analogous of the cylindricity condition. In particular, an extra-dimensional relic dependence in fields equation leads to a four-dimensional theory that should reveal the existence of the adjoint extra-space. Such a problem is typically bypassed by splitting the action of the theory in place of the corresponding motion equations; in fact, according to this point of view, we eliminate the extra-space dependence carrying out the extra-dimensional integral contained in the action.\\ 
This standard procedure, thus, would introduce an unnatural distinction between a formulation based on fields equation and one based on a variational principle, which instead find only in the fields equations its physical ground. In our work we want to reconcile both this formulations by showing how an integration on extra-coordinates has to be performed  even on fields equations; this is a consequence of the extra-dimensions compactification and of the way a four-dimensional observer reveals the dependence of a physical quantity on them.\\
Analogous considerations lead us to identify the gauge group with the translational one, along the Killing vectors in the extra-dimensional space, reproducing the correct gauge transformation laws either for gauge bosons and matter fields; in the latter case, we introduce them in such a way it is possible to geometrize gauge connections for spinors and to predict the associated charges conservation \cite{M04}.\\
In particular, in the first section we present the canonical approach based on the splitting of the action and show how the Einstein-Hilbert-Yang-Mills action (plus a cosmological term related to the curvature of the extra-dimensional space) can be derived; then, in view of reproducing non-Abelian gauge transformations on gauge bosons, we give same remarks about the link between four-dimensional and n-dimensional theories; in the following two sections, 3 and 4, from this analysis, we get the four-dimensional Einstein equations in presence of gauge bosons and Yang-Mills equations starting from the multidimensional Einstein-Hilbert ones; then, we reproduce gauge transformations for matter fields, by extra-dimensional translations, and the Dirac equation with gauge connections from its free multidimensional version is obtained. In section 5 brief concluding remarks follow.

\section{Geometrization of a gauge theory bosonic component in the canonical approach}
Let us consider a n-dimensional space-time manifold consisting of the direct sum 
of a four-dimensional one (the ordinary four dimensional space-time) and of a 
compact homogeneous one, $B^{K}$, whose Killing vectors $\xi_{\bar{P}}^{m}$ reproduce the algebra of the considered gauge group associated to the structure constants 
$C^{\bar{P}}_{\bar{N}\bar{M}}$.\\
We develop a theory in which general diffeomorphism covariance is broken explicitly, while the invariance under four-dimensional general coordinate transformations, as well as 
under translations along the Killing vectors, is maintained. According to Kaluza-Klein approach, we take n-bein vector fields of the form 
 \begin{equation}
\left\{\begin{array}{c} g_{\mu\nu}=\eta_{(\mu)(\nu)}e_{\mu}^{(\mu)}e_{\nu}^{(\nu)}\\
e_{\mu}^{(m)}=e^{(m)}_{m}\xi^{m}_{\bar{M}}A^{\bar{M}}_{\mu}\\
e_{m}^{(\mu)}=0\\\gamma_{mn}=\eta_{(m)(n)}e_{m}^{(m)}e_{n}^{(n)}\end{array}\right.,\quad \mu,\nu=0,..,3\quad m,n=4,..,n\label{nbein}
\end{equation}
where $A_{\mu}^{\bar{N}}$ denotes gauge bosons fields, while $g_{\mu\nu}$ and $\gamma_{mn}$, four and extra-dimensional metric, respectively. Furthermore, the dynamics of the theory is governed by the multidimensional Einstein-Hilbert action
\begin{equation}
S=-\frac{c^{3}}{16\pi G_{n}}\int_{V^{4}\otimes B^{K}} \sqrt{-j}{}^n\!R d^{4}xd^{K}y.
\end{equation}
As above mentioned, the dimensional reduction, in order to reproduce four-dimensional theories, cannot predict any dependence on extra-coordinates; in this sense, a classical approach is to split the action by carrying on the integration along the extra-dimensional space contained in its definition.\\ 
Therefore, by rewriting the scalar curvature in terms of $g_{\mu\nu}$, $A_{\mu}^{\bar{M}}$ and $\xi_{\bar{P}}^{m}$
we get the four-dimensional Einstein-Hilbert-Yang-Mills action including a cosmological constant contribution, i.e.
\begin{equation}
S=-\frac{c^{3}}{16\pi G}\int_{V^{4}}
\sqrt{-g}\bigg[R+\frac{1}{4}F^{\bar{M}}_{\mu\nu}F^{\bar{M}}_{\rho\sigma}g^{\mu\rho}g^{\nu\sigma}+R_{N}\bigg]d^{4}x;\label{az}
\end{equation}
the splitting procedure is performed by taking into account the following two assumptions
\begin{eqnarray}
\frac{1}{V}\int_{B^{K}}\sqrt{-\gamma}\gamma_{rs}\xi^{r}_{\bar{M}}\xi^{s}_{\bar{N}}d^{k}y=\eta_{\bar{M}\bar{N}}\qquad\qquad\qquad\label{3}\\
\frac{1}{V}\int_{B^{K}}\sqrt{-\gamma}\gamma_{rs}\xi^{r}_{\bar{M}}\xi^{s}_{\bar{N}}\gamma^{mn}\xi_{m}^{\bar{P}}\xi_{n}^{\bar{Q}}d^{k}y=\eta_{\bar{M}\bar{N}}\eta^{\bar{P}\bar{Q}}.\label{4}
\end{eqnarray}
However, this model is not completely satisfying: first we have that $A^{\bar{M}}_{\mu}$ transform (under extra-dimensional translations) as Abelian gauge bosons, so the approach cannot be extended to equations of motion. In fact, by varying the n-dimensional action, we find fields equations which depend on extra-coordinates, as forbidden in a pure four-dimensional theory; thus, from these statements, it would follow that a formulation based on the action has to be preferred respect to a motion equations based one. It is just the aim of the next sections to show that the unobservability of extra-dimensions overcomes such a dichotomy.\\ 

First of all, let us consider gauge bosons fields: they are defined (\ref{nbein}) and have always group index saturated with Killing vectors; then, we find the following transformation law ($\alpha^{\bar{M}}$ are arbitrary four-dimensional functions)
\begin{equation}
{\xi'}_{\bar{M}}^{\phantom1 m}{A'}^{\bar{M}}_{\mu}={\xi'}_{\bar{M}}^{\phantom1 m}(A^{\bar{M}}_{\mu}-\partial_{\mu}\alpha^{\bar{M}})=\xi_{\bar{M}}^{m}(A^{\bar{M}}_{\mu}+C^{\bar{M}}_{\bar{P}\bar{Q}}A^{\bar{P}}_{\mu}\alpha^{\bar{Q}}-\partial_{\mu}\alpha^{\bar{M}})\label{gbos}
\end{equation}
the first equality shows that $A^{\bar{M}}_{\mu}$ behave like Abelian gauge bosons, thus the geometrization would be possible only for Abelian gauge theories. However, a four-dimensional observer can see neither $\xi^{m}_{\bar{M}}$ nor their change and, in his point of view, the right equations to be addressed correspond to the last equality in (\ref{gbos}); in particular, if we integrate it on $B^{K}$, i.e. 
\begin{eqnarray*}
\frac{1}{V}\int_{B^{K}}\sqrt{-\gamma}{\xi'}_{\bar{M}}^{\phantom1 m}{A'}^{\bar{M}}_{\mu}d^{k}y=\frac{1}{V}\int_{B^{K}}\sqrt{-\gamma}\xi_{\bar{M}}^{m}(A^{\bar{M}}_{\mu}+\\+C^{\bar{M}}_{\bar{P}\bar{Q}}A^{\bar{P}}_{\mu}\alpha^{\bar{Q}}-\partial_{\mu}\alpha^{\bar{M}})d^{k}y\qquad\qquad\qquad\qquad\qquad\quad
\end{eqnarray*}
we find 
\begin{equation}
{A'}^{\bar{M}}_{\mu}=A^{\bar{M}}_{\mu}+C^{\bar{M}}_{\bar{P}\bar{Q}}A^{\bar{P}}_{\mu}\alpha^{\bar{Q}}-\partial_{\mu}\alpha^{\bar{M}};
\end{equation}
therefore we reproduce, in the four-dimensional theory, the correct transformation law for gauge bosons, even in the non-Abelian case.\\
In this way, we show how, to reproduce appropriate four-dimensional theories, we have to consider the relation between the multidimensional model and the four-dimensional observer physics, i.e. we need an integration on the extra-dimensional space, because it is unobservable at energy scale much less than the compactification scale.

\section{Four-dimensional observers}
All the multidimensional theories have to account for a four-dimensional phenomenology, thus if extra-dimensions exist they have to be unobserved at the present energy scales.\\ 
The compactification, together with quantum uncertainty 
\begin{equation}
\Delta x=\frac{\hbar c}{E},
\end{equation}
can explain why we do not perceive other dimensions and so their only physical effect is to include additional degrees of freedom in the dynamics.\\
When we reproduce a four-dimensional theory from a multidimensional one (including models with non-compactified extra-dimensions \cite{Wes90} \cite{Wes96}) we have to account for this unobservability, and any dependence of a physical law on extra-coordinates has to be eliminated.\\ 
As shown in the previous section, this issue is easily achieved by using a variational principle, since the action already contains an integration on the space-time coordinates.\\
However, in a formulation based on motion equations the problem of the unobservability still exists. To solve it, we have to consider the way how a four-dimensional observer reveals extra-dimensions; in fact he regards two points, whose coordinates differ only for their extra-dimensional labels, as the same point and thus an eigenstate of the four-dimensional position can be taken as 
\begin{equation}
|x\rangle = \frac{1}{V}\int dy |x;y\rangle,
\end{equation}
where V is the volume of $B^{K}$; in the latter, we consider a constant unit weight in taking the average value because of the extradimensional space homogeneity.
Therefore, the result of a measurement of a physical observable ($O$) on such an eigenstate is
\begin{equation}
\overline{O}=\frac{\left\langle x |O|x \right\rangle}{\left\langle x|x \right\rangle}=\frac{1}{V} \int O(x;y) dy.
\end{equation}

At the end, we see how an integration over extra-dimensions is required also in the motion equation approach, as a consequence of the extra-space unobservability.\\ 
The justification of such quantum point of view stands in the fact that the application of the model to physical interactions, such as the electro-weak one \cite{Io}, leads to predict a scale of compactification one order of magnitude greater than Planck's length; thus, we expect that geometric quantum phenomena arise in extra-dimensional physics.
   
\section{Fields equations approach}
The results of the previous section allow us to reproduce the four-dimensional theory starting from the motion equations directly; in fact, by varying the action we get Einstein vacuum equations in n-dimensions, i.e.
\begin{equation}
{}^n\!G_{AB}=0 \Rightarrow{}^n\!R_{AB}=0.
\end{equation}
Let us now split ${}^n\!R_{\mu\nu}$
\begin{eqnarray*}
{}^n\!R_{\mu\nu}=e_{\mu}^{(\mu)}e_{\nu}^{(\nu)}{}^n\!R_{(\mu)(\nu)}+e_{\mu}^{(m)}e_{\nu}^{(\nu)}{}^n\!R_{(m)(\nu)}+\quad\\+e_{\mu}^{(\mu)}e_{n}^{(\nu)}{}^n\!R_{(\mu)(n)}+
e_{m}^{(\mu)}e_{\nu}^{(n)}{}^n\!R_{(m)(n)}=e_{\mu}^{(\mu)}e_{\nu}^{(\nu)}{}^n\!R_{(\mu)(\nu)}
\end{eqnarray*}
where we used $R_{\mu m}=0$ and $R_{mn}=0$; then, we write $R_{(\mu)(\nu)}$ in terms of Riemann tensor's components 
\begin{equation}
{}^n\!R_{(\mu)(\nu)}=\eta^{(\rho)(\sigma)}{}^n\!R_{(\mu)(\rho)(\nu)(\sigma)}+\eta^{(m)(n)}{}^n\!R_{(\mu)(m)(\nu)(n)}\\
\label{Rmunu}
\end{equation}
and we calculate them from the relation by Ricci rotation coefficients $R_{(a)(b)(c)}$, i.e.
\begin{eqnarray}
{}^n\!R_{(a)(b)(c)(d)}=\partial_{(d)}R_{(a)(b)(c)}-\partial_{(c)}R_{(a)(b)(d)}+R_{(a)(f)(c)}\cdot\nonumber\\R^{(f)}_{(b)(d)}+R_{(a)(b)(f)}(R^{(f)}_{(c)(d)}-R^{(f)}_{(d)(c)})-R_{(a)(f)(d)}R^{(f)}_{(b)(c)}.\label{riem}
\end{eqnarray}
These coefficients, which we get from the n-bein (\ref{nbein}) by using
\begin{equation}
R_{(a)(b)(c)}=\frac{1}{2}(\lambda_{(a)(b)(c)}+\lambda_{(b)(c)(a)}-\lambda_{(c)(a)(b)}),
\end{equation}
being
\begin{equation}
\lambda_{(a)(b)(c)}=e_{(b)}^{b}e_{(c)}^{c}(\partial_{c}e_{(a)b}-\partial_{b}e_{(a)c}),
\end{equation}
have been implicitly calculated also in the splitting of the multidimensional Einstein-Hilbert action, obtaining the following expressions
\begin{equation}
\left\{\begin{array}{c}R_{(\mu)(\nu)(m)}=-\frac{1}{2}e^{\mu}_{(\mu)}e^{\nu}_{(\nu)}e_{(m)r}\xi^{r}_{\bar{M}}F^{\bar{M}}_{\mu\nu}\\\\
R_{(m)(\mu)(\nu)}=\frac{1}{2}e^{\mu}_{(\mu)}e^{\nu}_{(\nu)}e_{(m)r}\xi^{r}_{\bar{M}}F^{\bar{M}}_{\mu\nu}\\\\
R_{(\mu)(m)(n)}=-2e_{s\{(m)}e^{n}_{(n)\}}e^{\mu}_{(\mu)}
A^{\bar{M}}_{\mu}\xi^{s}_{\bar{P}}
\xi^{\bar{Q}}_{n}C^{\bar{P}}_{\bar{Q}\bar{M}}\\\\
R_{\bar{m}\bar{n}\bar{\mu}}=-2e_{s[(m)}e^{n}_{(n)]}e^{\mu}_{(\mu)}
A^{\bar{M}}_{\mu}\xi^{s}_{\bar{P}}
\xi^{\bar{Q}}_{n}C^{\bar{P}}_{\bar{Q}\bar{M}}\\\\
R_{(m)(n)(p)}=\frac{1}{2}(e_{(m)s}
e^{n}_{(n)}e^{p}_{(p)}+e_{(n)s}
e^{n}_{(p)}e^{p}_{(m)}-\\\\e_{(p)s}
e^{n}_{(m)}e^{p}_{(n)})C^{\bar{P}}_{\bar{Q}\bar{M}}
\xi^{s}_{\bar{P}}\xi^{\bar{Q}}_{p}\xi^{\bar{M}}_{n}.
\end{array}\right.\label{riccicoe}
\end{equation}
Now, we rewrite the equation ($\ref{Rmunu}$) with the help of the relations (\ref{riem}) (\ref{riccicoe}) and, after some algebra, we find 
\begin{eqnarray*}
{}^n\!R_{(\mu)(\nu)}=R_{(\mu)(\nu)}-\frac{1}{4}\gamma_{rs}\xi^{r}_{\bar{M}}\xi^{s}_{\bar{N}}e_{(\mu)}^{\mu}e_{(\nu)}^{\nu}F^{\bar{M}}_{\mu\rho}F^{\bar{N}\rho}_{\nu}+\qquad\quad\\+e_{(\mu)}^{\mu}e_{(\nu)}^{\nu}A_{\mu}^{\bar{M}}A_{\nu}^{\bar{N}}(C^{\bar{P}}_{\bar{Q}\bar{M}}C^{\bar{Q}}_{\bar{P}\bar{N}}+C^{\bar{P}}_{\bar{Q}\bar{M}}C^{\bar{R}}_{\bar{S}\bar{N}}\gamma_{rs}\xi^{r}_{\bar{P}}\xi^{s}_{\bar{R}}\gamma^{tu}\xi_{t}^{\bar{Q}}\xi_{u}^{\bar{S}});
\end{eqnarray*}
then, we obtain $G_{\mu\nu}$ and we can carry on the integration along $B^{K}$, getting (by using $R=0$, (\ref{3}) and (\ref{4})) the splitting 
\begin{eqnarray}
\frac{1}{V}\int_{B^{K}}{}^n\!G_{\mu\nu}\sqrt{\gamma}d^{k}y =G_{\mu\nu}+\frac{1}{2} F^{\bar{M}}_{\mu\rho}F^{\bar{M}\phantom1\rho}_{\nu}-\\
-\frac{1}{8}g_{\mu\nu} F^{\bar{M}}_{\rho\sigma}F^{\bar{N}\rho\sigma}-\frac{1}{2}g_{\mu\nu}R_{N}=0\qquad\qquad\quad\nonumber
\end{eqnarray}
i.e. Einstein equations in presence of bosons fields (coupling constants are contained in the fields $A^{\bar{M}}_{\mu}$).\\
In the same way from ${}^n\!R_{m\mu}=0$ we obtain, after averaging on extra-dimensions, Yang-Mills equations
\begin{equation}
\frac{1}{V}\int_{B^{K}} \xi^{m}_{\bar{M}}R_{m\mu}\sqrt{\gamma}d^{k}y=\frac{1}{2}g^{\rho\sigma}(\partial_{\sigma}F^{\bar{M}}_{\rho\mu}-C^{\bar{M}}_{\bar{R}\bar{S}}F^{\bar{S}}_{\rho\mu}A^{\bar{R}}_{\sigma})=0.
\end{equation}
Finally, from ${}^n\!R_{mn}=0$, we get the relations
\begin{eqnarray}
\nonumber\xi^{m}_{\bar{M}}\xi^{n}_{\bar{N}}R_{mn}=F^{\bar{M}}_{\mu\nu}F^{\bar{N}\mu\nu}+\frac{1}{4}C^{\bar{P}}_{\bar{Q}\bar{V}}C^{\bar{R}}_{\bar{S}\bar{T}}\frac{1}{V}\cdot\qquad\\\cdot\int\sqrt{\gamma}\gamma_{mp}\xi^{m}_{\bar{M}}\xi^{p}_{\bar{P}}\gamma_{rn}\xi^{r}_{\bar{R}}\xi^{n}_{\bar{N}}\gamma^{vt}\xi_{v}^{\bar{V}}\xi_{t}^{\bar{T}}\gamma^{qs}\xi_{q}^{\bar{Q}}\xi_{s}^{\bar{S}}d^{k}y
\end{eqnarray}
and it is easy to realize they are not compatible with a real Yang-Mills theory; however we infer that the latter equations would just provide the dynamics of some fields contained in the extra-dimensional metric, which here we have taken as constants (as well known for the 5-dimensional theory). In order to reproduce a gauge theory we can set these fields in such a way that they just introduce a dependence of physical quantities on four-dimensional variables \cite{Io2}, according to Dirac large number hypothesis \cite{Dir}; the discussion of such fields dynamics would not add any additional physical insight in the present analysis.\\
We remark that, unless taking the integration along $B^{K}$, we cannot eliminate the dependence on extra-dimensional variables and reproduce the four-dimensional theory; thus the model would have an unnatural preference for a formulation based on the splitting of the action. Moreover, we predict that, at energy scales such that extra-dimensions become observable, we must modify ordinary motion equations and extra-coordinates are involved in the dynamics.

\section{Coupling with matter fields}
Despite the issue of geometrization, we have to introduce spinors even in Kaluza-Klein theories, in order to reproduce matter fields involved in the Standard Model. To reproduce the conservation of gauge charges (from the invariance of the theory under extra-dimensional translations) and the right coupling with bosons fields, we can assume a suitable dependence on extra-coordinates for the spinor fields \cite{M04}\cite{Io2}.\\ 
In particular, we address the following extra-dimensional dependence
\begin{equation}
\Psi_{r}=\frac{1}{V}e^{-iT_{\bar{P}rs}\lambda_{\bar{Q}}^{\bar{P}}\Theta^{\bar{Q}}(y)}\psi_{s}(x)
\end{equation}
where $T_{\bar{M}}$ are the gauge generators, $\Theta(y)$ functions on extra-coordinates compatible with the symmetries of $B^{K}$ and for the constant matrix $\lambda$ we have 
\begin{equation}
\label{lam}(\lambda^{-1})^{\bar{P}}_{\bar{Q}}=\frac{1}{V}\int_{B^{K}}
\sqrt{-\gamma}\Big(\xi^{m}_{\bar{Q}}\partial_{m}\Theta^{\bar{P}}(y)\Big)d^{K}y.
\end{equation}
We emphasize that $\Theta^{\bar{N}}$ cannot be scalar functions, otherwise the expression (\ref{lam}) vanishes; a natural form for them, invariant under isometries on $B^{K}$, is a scalar density of weight $\frac{1}{2}$,  i.e. $\Theta^{P}=\sqrt{\gamma}\phi^{P}(y^{l})$, being $\phi^{P}$ non-constant scalars.\\
If we consider translations along $B^{K}$, we find that their transformation law reads
\begin{equation}
\psi'_{r}=(\psi_{r}+i\alpha^{\bar{Q}}\lambda^{\bar{N}}_{\bar{Q}}
\xi^{m}_{\bar{N}}\partial_{m}\Theta^{\bar{P}}(y)T_{\bar{P}rs}\psi_{s})
\end{equation}
and so, after averaging on extra-dimensions, we get for a four-dimensional observer       
\begin{equation}
\psi'_{r}=\psi_{r}+i\alpha^{\bar{Q}}T_{\bar{Q}rs}\psi_{s}
\end{equation}
which reproduce the right transformation for matter fields.\\
Therefore, extra-dimensional translations, together with the unobservability of $B^{K}$, induce gauge transformations into four-dimensional theories.\\
Starting from the multidimensional free spinor action, we get the expected generalization of the Dirac equation, i.e.
\begin{equation}
\gamma^{(A)}e_{(A)}^{A}(\partial_{A}+\Gamma_{A})\Psi=0\label{Direq}
\end{equation}
where $\gamma^{(A)}$ are n Dirac matrices and $\Gamma_{A}$ spinorial connections. As consequence of the explicit breaking for the multidimensional Lorentz group, we take the standard form for four-dimensional quantities $\Gamma_{(\mu)}$ \cite{B82} 
\begin{equation}
\Gamma_{(\mu)}=-\frac{1}{4}\gamma^{(\nu)}\nabla_{(\mu)}\gamma_{(\nu)}
\end{equation}
(the quantities are all four-dimensional)
and we choose a particular form for extra-dimensional ones (required by the consistency of the action's splitting with the four-dimensional Einstein-Yang-Mills-Dirac lagrangian \cite{Io}\cite{Io2})
\begin{equation}
\label{spcon}\Gamma_{(m)}=\frac{i}{V^{K}}\lambda^{\bar{P}}_{\bar{Q}}T_{\bar{P}}\int_{B^{K}}
\sqrt{-\gamma}e^{m}_{(m)}\partial_{m}\Theta^{\bar{Q}}d^{K}y.
\end{equation}
Now, via the dimensional reduction of the equation (\ref{Direq}), we have
\begin{eqnarray*}
\gamma^{(\mu)}(e_{(\mu)}^{\mu}\partial_{\mu}+e_{(\mu)}^{m}\partial_{m}+\Gamma_{(\mu)})\Psi+\gamma^{(m)}(e_{(m)}^{m}\partial_{m}+\Gamma_{(m)})\Psi=\\=
e^{-iT_{\bar{P}}\lambda_{\bar{Q}}^{\bar{P}}\Theta^{\bar{Q}}}[\gamma^{(\mu)}(e_{(\mu)}^{\mu}\partial_{\mu}+e_{(\mu)}^{\mu}iA_{\mu}^{\bar{M}}
\xi_{\bar{M}}^{m}T_{\bar{P}}\lambda^{\bar{P}}_{\bar{Q}}\partial_{m}\Theta^{\bar{Q}}+\\+\Gamma_{(\mu)})+\gamma^{(m)}(-ie_{(m)}^{m}T_{\bar{P}}\lambda^{\bar{P}}_{\bar{Q}}\partial_{m}\Theta^{\bar{Q}}+\Gamma_{(m)})]\psi=0\qquad
\end{eqnarray*}
which can be rewritten as 
\begin{eqnarray*}
[\gamma^{(\mu)}(e_{(\mu)}^{\mu}\partial_{\mu}+e_{(\mu)}^{\mu}iA_{\mu}^{\bar{M}}
\xi_{\bar{M}}^{m}T_{\bar{P}}\lambda^{\bar{P}}_{\bar{Q}}\partial_{m}\Theta^{\bar{Q}}+\Gamma_{(\mu)})+\\+\gamma^{(m)}(-ie_{(m)}^{m}T_{\bar{P}}\lambda^{\bar{P}}_{\bar{Q}}\partial_{m}\Theta^{\bar{Q}}+\Gamma_{(m)})]\psi=0;
\end{eqnarray*}
integrating on extra-dimensions and using (\ref{lam}) (\ref{spcon}), we arrived to the four-dimensional Dirac equation with gauge couplings
\begin{equation}
\gamma^{(\mu)}(e_{(\mu)}^{\mu}\partial_{\mu}+e_{(\mu)}^{\mu}iA_{\mu}^{\bar{M}}
T_{\bar{M}}+\Gamma_{(\mu)})\psi=0.
\end{equation}
Thus, we see how the averaging procedure, required by the extra-dimensions unobservability, makes account for the appearance of gauge bosons in the Dirac equation, as viewed by a four-dimensional observer. Again, we stress how it is just the impossibility to reveal the extra-dimensional space which allow to implement extra-dimensional coordinates transformations as gauge symmetries of the four-dimensional theory, either for bosons as well as for matter fields. 

\section{Concluding remarks}
In this paper, we have shown how to reconcile Kaluza-Klein models for arbitrary compact homogeneous spaces with gauge theories in the ordinary four-dimensional space-time, so increasing the physical plausibility of this approach and of extra-dimensions, in general.\\
In order to achieve this issue, the four-dimensional theory is regarded as a phenomenological limit of the fundamental n-dimensional model, based on the unobservability of extra-dimensions.\\
We infer that analogous considerations should be made for any theory with unobservable extra-dimensions, even if non-compact.\\
Moreover, in this scenario the difference between gravity and other interactions arise because they are different metric components. In particular, since no cosmological gauge boson's fields are observed, by assuming a vacuum state with vanishing mist metric components (linking $V^{4}$ and $B^{K}$), we can treat the gauge carriers as perturbations to the cosmological background and quantize them in a perturbative scheme. Addressing to this point of view, the coexistence, in modern physics, of a classical formulation for gravity and gauge fields (which like in the case of Yang-Mills bosons manifest their quantum behavior) would be explained. The main issue of the present analysis has to be regarded how the multidimensional Einstein equations plus the Dirac one for spinor fields reproduce the four-dimensional Einstein-Yang-Mills-Dirac dynamics, under dimensional reduction. The key feature of our approach is to require an averaging procedure on the extra-dimensions, which comes out from the impossibility of detecting the extra-space $B^{K}$ by a four-dimensional observer. The average on extra-dimensions is taken with uniform weight, in view of the $B^{K}$ homogeneity, and allows to reduce the dynamics to a pure four-dimensional one, i.e. each physical observable depends only on usual four coordinates.\\
By other words, we stress how to obtain gauge symmetries, within a Kaluza-Klein approach, is not enough to break down the full Lorentz group, but it is also necessary to require that the extra-space is not detectable on the energy scale accessible to the observer. In fact, in our field equation approach, to reveal extra-dimensions would be equivalent to recognize that no real gauge bosons exist, but simply extra-dimensional metric components.\\

\section{Acknowledgments}  
Erica Cerasti is thanked for her help in upgrading the English of the manuscript.


\begin{thebibliography}{}

\bibitem{1}
T. Kaluza, \emph{On the Unity Problem of Physics}, Sitzungseber. Press. Akad. Wiss. Phys. Math. Klasse {\bf 966}, (1921)

\bibitem{2}
O.Klein, Z. F. Physik {\bf37}, (1926), 895
  
\bibitem{3}
O. Klein, Nature {\bf118}, (1926), 516 

\bibitem{Ke68}
R. Kerner,
{\it Ann. Inst. Henri Poincaré} {\bf 9}, (1968), 143


\bibitem{MKKT}
T.Appelquist, A.Chodos, P.Frund, \emph{Modern Kaluza-Klein theories}, Addison Wesley Publishing, Inc., (1987) 

\bibitem{OW97}
I.M. Overduin and PS. Wesson,
{\it Physics Reports},  {\bf 283}, (1997), 303

\bibitem{W81}
E. Witten, {\it Nucl. Phys.}, {\bf B186}, (1981), 412

\bibitem{W83}
C. Wetterich, {\it Nucl. Phys.} {\bf B222}, (1983), 20

\bibitem{Lan}
L. Landau, E.M. Lifsits,
{\it Teoria dei campi}, Editori Riuniti, (1999)

\bibitem{M04}
G. Montani, 
{\it Int. J. Mod. Phys.}, {\bf 44}, (2005), 653

\bibitem{Wes90}
P.S. Wesson,
{\it Gen. Rel. Grav.}, {\bf 22}, (1990), 707

\bibitem{Wes96}
P.S. Wesson et al.,
{\it Int. J. Mod. Phys.}, {\bf A11}, (1996), 3247

\bibitem{Io}
F. Cianfrani, G. Montani,
in preparation


\bibitem{Io2}
F. Cianfrani, A. Marrocco, G. Montani,
{\it Int. Journ. Mod. Phys D}, in press

\bibitem{Dir}
P. M. Dirac, {\it The variation of G and quantum
theory}, proc. of the M.Grossmann meet. on Gen.
Rel., R. Ruffini Editor, North Holland Pub. Ed.,
(1982)

\bibitem{B82}
N.D. Birrel, P.C. Davis, \emph{Quantum field theory
in curved space-time}, Cambridge University press, Cambridge,
(1982)


\end{thebibliography}
\end{document}